\documentclass{appolb}

\usepackage{cite,graphicx,amsmath,amsfonts,amssymb,mathrsfs}

\newcommand{\etc}{{\it etc.}}
\newcommand{\eq}{Eq.}

\newcommand{\fig}{Figure}

\newcommand{\Ref}{Ref.}
\newcommand{\Refs}{Refs.}

\newcommand{\Tab}{Table}

\newcommand{\stheta}{\sin^22\theta_{13}}
\newcommand{\deltacp}{\delta_{\mathrm{CP}}}
\newcommand{\ldm}{\Delta m_{31}^2}
\newcommand{\sdm}{\Delta m_{21}^2}
\newcommand{\equ}[1]{\eq~(\ref{equ:#1})}
\newcommand{\figu}[1]{\fig~\ref{fig:#1}}

\newcommand{\bi}{\begin{itemize}}
\newcommand{\ei}{\end{itemize}}

\begin{document}
%\date{\today}
\pagestyle{plain}

\newcount\eLiNe\eLiNe=\inputlineno\advance\eLiNe by -1
\title{FUTURE PRECISION NEUTRINO EXPERIMENTS \\
AND THEIR THEORETICAL IMPLICATIONS%
}
\author{Walter WINTER\thanks{E-mail: {\tt winter@physik.uni-wuerzburg.de}}%
\address{Institut f{\"u}r theoretische Physik und Astrophysik, Universit{\"a}t W{\"u}rzburg \\
Am Hubland, D-97074 W{\"u}rzburg, Germany}}
\maketitle

\begin{abstract}
Future neutrino oscillation measurement focus, at first priority, on the discovery
of $\stheta$ describing the coupling between solar and atmospheric oscillations.
In this talk, we briefly discuss the prospects to measure $\stheta$ by future reactor
and beam experiments, and we illustrate the usefulness of these measurements (and their
precision) from the 
theoretical point of view with an example.
\end{abstract}

\section{Introduction}

Recent neutrino oscillation measurements have established two different
mass squared differences ($\sdm$ and $\ldm$) and two large mixing angles ($\theta_{12}$ and
$\theta_{23}$) in the lepton sector, 
which describe the solar and atmospheric neutrino oscillations
(see, \eg, \Ref~\cite{Schwetz:2006dh}). While the atmospheric mixing angle $\theta_{23}$ is large,
and may even be maximal, maximal mixing is excluded for the solar angle $\theta_{12}$.
The third mixing angle $\theta_{13}$, which may couple the solar and atmospheric
oscillations, is however unknown, and there exists and upper bound $\stheta \lesssim 0.1$
mainly coming from the Chooz reactor experiment~\cite{Apollonio:1999ae}.
This mixing angle is the key to sub-leading effects, such as the determination
of the neutrino mass hierarchy, or the measurement of the leptonic Dirac CP phase.
Therefore, it will be the first priority of upcoming neutrino oscillation experiments
to establish $\stheta > 0$. For a recent review on neutrino oscillation physics,
see, \eg, \Ref~\cite{Strumia:2006db}.

From the theoretical point of view, neutrino mass models can be described by textures,
GUTs, anarchy arguments, flavor symmetries, \etc. In order to test specific models, some 
experimental observables are very well suited as performance indicators,
such as the magnitude of $\theta_{13}$ (see, \eg, \Tab~1 in \Ref~\cite{Anderson:2004pk}),
the neutrino mass hierarchy (see, \eg, \Ref~\cite{Albright:2006cw}), and
deviations from maximal atmospheric mixing (see, \eg, \Ref~\cite{Antusch:2004yx}).
This can be easily understood in terms of symmetries, which may force $\theta_{13}$ to $0$
or $\theta_{23}$ to $\pi/4$. In addition, the normal and inverted hierarchy mass schemes
lead to very different structures of the neutrino mass matrix. Though these performance
indicators are well-suited for models describing the lepton sector, the usual approach
to parameterize $U_{\mathrm{PMNS}}$ in the same way as $V_{\mathrm{CKM}}$ (and to compare
the magnitudes of the mixing angles) motivates performance indicators indicative for
quark-lepton unification as well. One such indication might be 
\begin{equation}
\pi/4 \simeq \theta_{23} \simeq \theta_{12} + \theta_C \, , \label{equ:qlc} 
\end{equation}
which is often referred to as ``quark-lepton complementarity''~\cite{qlc} (QLC).

In this talk, we focus on two main questions: What are the upcoming possibilities
to obtain precision measurements on $\stheta$ (and sub-leading effects), and
what precision is needed from the theoretical point of view? 
For the latter question, we will show a specific
example based on quark-lepton complementarity assumptions.

\section{Future precision neutrino measurements}

Reactor neutrino experiments are probably the next class of experiments providing more
stringent limits on $\theta_{13}$. Similar to the Chooz experiment, a detector is placed
in a distance $\simeq 1 \sim 2 \, \mathrm{km}$ from the reactor core(s), while a 
(preferably identical) near-detector measures the flux in a distance of a few hundred meters and reduces systematical errors~\cite{Minakata:2002jv,Huber:2003pm,Anderson:2004pk}. The oscillation probability 
is then, to first approximation, given by the very simple formula
\begin{equation}
1-P_{\bar{e} \bar{e}} \simeq \stheta \, \sin^2 \frac{\ldm L}{4E} \, ,
\end{equation}
which allows for a very clean determination of $\stheta$. Of course, measuring the
difference to the ``1'' implies that systematics is crucial for this type of experiment,
in particular, for a large luminosity setup~\cite{Huber:2003pm}. Two examples for future
potential reactor experiments are Double Chooz~\cite{Ardellier:2004ui} and Daya Bay~\cite{Guo:2007ug},
where the first one operates in the lower luminosity regime and is likely to be statistics limited, and the second will be more sensitive to systematics while testing smaller values of $\theta_{13}$.

Compared to reactor experiments, neutrino beam experiments use the $\nu_e \leftrightarrow \nu_\mu$
oscillation channel, which is very sensitive to $\stheta$, the neutrino mass hierarchy, and leptonic CP violation. However, the information has to be extracted from a complicated convolution of parameters.
Even if both the CP-conjugated neutrino and antineutrino channels are used, three discrete degeneracies
are remaining, leading to an overall $2^3=8$-fold degeneracy~\cite{Fogli:1996pv
%,Minakata:2001qm,Burguet-Castell:2001ez,Barger:2001yr
}. In addition, correlations among the
oscillation parameters spoil the measurements~\cite{Huber:2002mx}. Therefore, strategies to resolve these degeneracies have been discussed in the past few years, which depend on the experiment class considered. Compared to the reactor experiments, beam experiments are not only sensitive to $\theta_{13}$, but also to the sub-leading effects. In addition, they can, depending on the experiment class, go to much smaller values of $\theta_{13}$. However, the dependence on the sub-leading effects constrains the exclusion power with respect to $\theta_{13}$. In that sense, the beam and reactor experiment classes are very complementary physics
concepts.

\begin{figure}[t]
\begin{center}
\includegraphics[width=7cm]{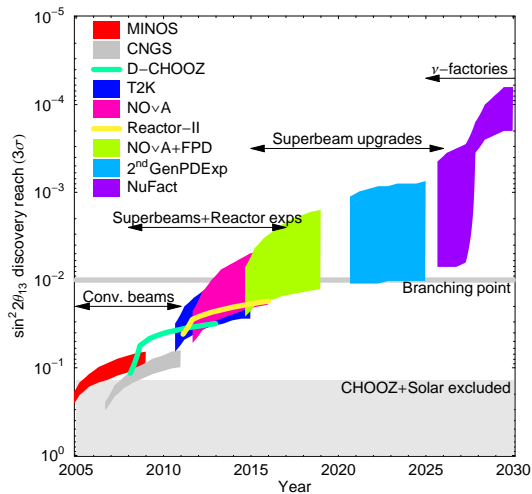}
\vspace*{-0.5cm}
\end{center}
\caption{\label{fig:disclimits} Possible evolution of the $\stheta$ discovery potential as function of time. The band reflect the unknown true value of $\deltacp$. The label ``FPD'' stands for ``Fermilab Proton Driver'', and ``$2^\mathrm{nd}$GenPDExp'' for a second-generation superbeam upgrade using a proton driver.
Figure from \Ref~\cite{Albrow:2005kw}.}
\end{figure}

A currently running neutrino beam experiment is the MINOS experiment at Fermilab~\cite{Michael:2006rx}, which uses a muon neutrino
beam obtained from pion (and kaon) decays. The T2K~\cite{Itow:2001ee} and NO$\nu$A~\cite{Ayres:2004js} superbeams using higher target powers, bigger detectors, and the off-axis technology to improve the signal over background ratio
are future potential neutrino beam experiments with very good discovery potentials of $\stheta$. Beyond
that, superbeam upgrades operated on- or off-axis (see, \eg, \Refs~\cite{Barger:2006kp
%,Barger:2007jq
} for a comparison of different options), beta beams using electron neutrino beams from radioactive isotope decays (see, \eg, \Refs~\cite{BurguetCastell:2005pa
%,Huber:2005jk
} for higher gamma options), and neutrino factories using muon decays for neutrino production~\cite{Geer:1997iz} are being discussed. We show in \figu{disclimits} a possible evolution of the $\stheta$ discovery potential as function of time. In this figure, the bands reflect the dependence on the unknown value of $\deltacp$ implemented by nature. One can clearly identify the different generations of experiments and the difference between reactor and beam experiments in this figure.

The study of a neutrino factory as a potential ultimate precision instrument has recently been drawing
attention. This type of experiment may discover $\stheta$ for values as small as $10^{-4}$ to $10^{-5}$ (see figure) and measure $\deltacp$ with a precision up to 10 degrees~\cite{Huber:2004gg} ($1\sigma$, $\stheta=10^{-3}$). An international design study including cost study is currently being launched. For the neutrino factory experiment, degeneracies may be resolved by using a second baseline at $L_2 \simeq 7 \, 500 \, \mathrm{km}$ in combination with a shorter baseline $L_1 \simeq 3 \, 000 \, \mathrm{km}$ to $4 \, 000 \, \mathrm{km}$~\cite{Huber:2006wb}. At this very long ``magic baseline''~\cite{Huber:2003ak}, matter effects make the dependence on $\deltacp$ disappear and allow for a clean measurement of $\theta_{13}$ -- and even the matter density~\cite{Minakata:2006am,Gandhi:2006gu}. Therefore, the combination of these two baselines is the configuration currently
discussed as the standard setup of a neutrino factory. Further potential improvements include the detection system~\cite{Anselmo} and the use of alternative channels~\cite{Huber:2006wb}.  In the following, we will discuss how such high precision measurements in the lepton sector can actually be motivated.

\section{Why do we need these measurements? - An example}

In order to test the impact of future precision measurements on theory space, we need to
have a procedure, which is as unbiased as possible, to create a parameter space of alternative 
equivalent theories. We use a bottom-up approach at the texture level in \Refs~\cite{Plentinger:2006nb,Plentinger:2007px}: We generate a sample of 
realizations compatible with current data, \ie, neutrino mass matrices including valid order one coefficients,
  from very generic assumptions. Then we identify
the leading orders to reconstruct the structure of the Yukawa couplings (texture). Compared a
conventional approach to generate the Yukawa coupling structure first, and then fit the order
one coefficients, our approach does not require diagonalization and allows for a simple inclusion
of the mass spectrum to reduce complexity. The objectives of this approach are two-fold: First, we
want to generate all (new) possibilities from a given set of generic assumptions. And second, we 
want to study the generated parameter space. A spin-off of the second objective is that we can
also discuss how experiments affect this parameter space.

We take our generic assumptions from the QLC hypothesis in \equ{qlc}: We assume that all mixing angles and
hierarchies be generated by powers of $\epsilon \simeq \theta_C$, including $\epsilon^0 = \mathcal{O}(1)$
(corresponding to maximal mixing for the mixing angles).
 It can be motivated by the observation 
that mixing angles $\sim \theta_\text{C}$ (and powers $\sim\theta_\text{C}^n$ thereof), 
as well as maximal mixing, can be readily obtained in models from flavor symmetries.
Consequently, the observed large leptonic mixing angles $\theta_{12}$
and $\theta_{23}$ can only arise as a result of taking the product of the charged
lepton mixing matrix $U_\ell$ and the neutrino mixing matrix
$U_\nu$ in the PMNS mixing matrix
\begin{equation}
U_{\rm PMNS}=U_\ell^\dagger U_\nu \, .
\label{equ:upmns}
\end{equation}
Simple conventional quark-lepton complementarity
implementations, such as $U_{\rm PMNS} \simeq V_{\rm CKM}^\dagger U_{\rm bimax}$, emerge as special cases
in this approach (see, \eg, \Refs~\cite{qlcbimax}), but are not the exclusive solutions. For example, the charged lepton sector may actually induce two large mixing angles. 
An implication of this type of generic assumptions is that the quark and lepton sectors must be
somehow related, such as by a remnant of a unified theory $\epsilon$. In addition, it is a good 
application to illustrate why it is useful to parameterize $U_{\rm PMNS}$ and $V_{\rm CKM}$ in the same way.

\begin{figure}[t]
\begin{center}
\includegraphics[width=\textwidth]{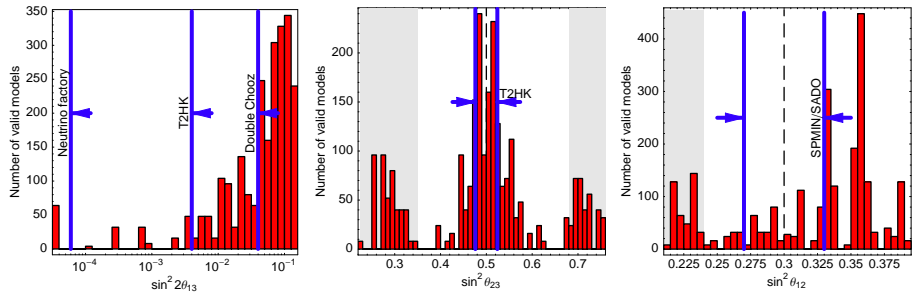}
\vspace*{-0.5cm}
\end{center}
\caption{\label{fig:histo} Distributions of mixing angles for our realizations in the effective $3 \times 3$ case, as well as projected precisions of future experiments. Figure taken from \Ref~\cite{Plentinger:2006nb}.}
\end{figure}

We show in \figu{histo} the distribution of mixing angles for all realizations we have generated from our generic assumptions. As far as the different mixing angles are concerned, there a three major observations: First, future measurements of $\stheta$ will strongly reduce this parameter space. Without a neutrino factory, too many possibilities different from $\stheta \simeq 0$ may be remaining. Second, since maximal mixing was used as one of the generic input assumptions, testing small deviations from maximal mixing is not very indicative for this class of models. And third, $\theta_{12}$ is a very good performance indicator
for QLC-based models, since the peak position in \figu{histo}, right, will be determined by the magnitude of $\theta_C$.

In \Ref~\cite{Plentinger:2006nb}, we do not only study the parameter space, but we also generate specific textures. In addition, we present in \Ref~\cite{Plentinger:2007px} an extension to the see-saw mechanism, leading to $1 \, 981$ valid texture sets for the normal hierarchy (for the case of small $\stheta$). Furthermore, we introduce complex phases in \Ref~\cite{Winter:2007yi}, connect specific textures to different distributions of observables, and illustrate that measuring $\deltacp$ with Cabibbo-angle precision $\sim 11^\circ$ is useful. This precision might be achieved by the neutrino factory experiment discussed above~\cite{Huber:2004gg}. The use of textures can be motivated as an intermediate result to be used for model building. For example, Froggatt-Nielsen models~\cite{Froggatt:1978nt} using flavor symmetries predict certain structures in the Yukawa couplings. Our textures can be used to study the parameter space of discrete flavor symmetries~\cite{Plentinger:prep}.

\section{Summary and conclusions}

It is the objective of future reactor and accelerator neutrino oscillation experiments
to probe neutrino masses and mixings to very high precisions. The primary goal is the determination of the
small mixing angle $\stheta$, because it is the key to the mass hierarchy and CP violation measurements.
But what motivates a high precision, such as obtained for a neutrino factory? 

In order to test the impact of precision, we have systematically generated a parameter space of mass matrices from generic assumptions.  These generic assumptions can be motivated by the fact that  the quark and lepton mixing matrices $U_{\rm PMNS}$ and $V_{\rm CKM}$ are usually parameterized in the same way, but show strong qualitative differences. We believe that such a comparison can only be justified in light of a possible connection. We have parameterized this connection in terms of the Cabibbo angle as a generic assumption, and we have illustrate that the precision at the level of a neutrino factory can be well-motivated. 

We conclude that even for the known three-flavor oscillation physics, the higher precision of future facilities will be very useful. Beyond that, the search for non-standard physics may motivate such a facility. Finally, a possible connection between $\deltacp$ and leptogensis makes the test of leptonic CP violation the ultimate goal.

\vspace{-1.5cm}

\renewcommand\refname{}

% small

%\bibliography{references}
%\bibliographystyle{apsrev}

\end{document}